\documentclass[a4paper,11pt]{article}
\usepackage{pos}
\usepackage{amsmath,mathtools,commath}
\usepackage{graphicx}
\usepackage{physics}
\usepackage{listings}
\usepackage{subcaption}
\usepackage{todonotes}
\usepackage{caption}
\usepackage{wrapfig}

\title{Spectral reconstruction techniques, their shortcomings and relevance to the electric conductivity coefficient}
\ShortTitle{Spectral reconstructions and the electric conductivity coefficient}

\author*[a]{C. Andratschke}
\author[a]{B. B. Brandt}
\author[b]{E. Garnacho-Velasco}
\author[a]{L. Pannullo}
\author[c]{S. Singh}
\author[d]{A. Dean M. Valois}

\affiliation[a]{Fakultät für Physik, Universität Bielefeld,\\
  Universitätsstraße 25, 33615 Bielefeld, Germany}

\affiliation[b]{Institute of Theoretical Physics and Astronomy, ELTE Eötvös Loránd University,\\
Pázmány P. sétány 1/A, H-1117 Budapest, Hungary}

\affiliation[c]{Helmholtz-Institut für Strahlen- und Kernphysik (HISKP), Universität Bonn,\\
Nussallee 14-16, 53115 Bonn, Germany}

\affiliation[d]{CAFPE and Departamento de Física Teórica y del Cosmos, Universidad de Granada,\\
E-18071 Granada, Spain}

\emailAdd{christian.andratschke@uni-bielefeld.de}

\abstract{Spectral reconstruction is a well-studied numerically ill-posed problem which arises due to the relation of the Euclidean correlator to the spectral function via an inhomogeneous Fredholm equation of the first kind. Several different methods are on the market to resolve this issue, each taking different approaches and assumptions. In these proceedings, we focus on implementing and testing a machine learning framework for spectral reconstruction, as well as implementing a novel method of estimating the behavior of the spectral function in the vicinity of vanishing frequency, which we denote as the multipoint method, and compare these methods to well-established spectral reconstruction techniques from the literature using mock data. As a physics application, we apply the reconstruction techniques to quenched lattice data for the correlation function in the vector channel at non-zero external magnetic field to extract the spectral function and the electric conductivity through its behaviour at vanishing frequency via a Kubo formula.}

\FullConference{The 42nd International Symposium on Lattice Field Theory (LATTICE2025)\\
2-8 November 2025\\
Tata Institute of Fundamental Research, Mumbai, India\\}

\begin{document}
\maketitle

\section{Introduction}
The numerical solution of spectral reconstruction problems is relevant in a wide variety of physical contexts, ranging from geophysics to astrophysics. In the high-energy physics community, particularly in the study of strong interactions, these methods allow one to obtain the spectral function from Euclidean correlators, which can be calculated non-perturbatively using lattice QCD simulations. At finite temperature, these spectral functions encode crucial information about the transport properties of the quark-gluon plasma, which existed in the early universe and can be currently studied in heavy-ion collisions. However, spectral reconstruction methods using lattice data face several issues. Most prominently, the reduced amount of available points for the reconstruction yields an ill-posed inverse problem.

 Over the years, a number of different reconstruction techniques have been applied to the inverse problem with varying degrees of success. One of the initially most widely used examples is the Maximum Entropy Method (MEM)~\cite{Bryan_MEM,Asakawa_2001}, which has been applied to several different problems, including the extraction of transport coefficients as well as meson spectral functions~\cite{Ding_2011, Meyer_2011, Ding_2019}. Other commonly used methods include the Gaussian method~\cite{Horak_2022}, the Hansen-Lupo-Tantalo (HLT) method~\cite{Hansen:2019idp} or the Backus-Gilbert (BG) method~\cite{BG1968}, among others. Recently, several neural networks have been proposed to study the inverse problem, e.g.~\cite{Wang:2023fhc, Kades:2019wtd}. Due to the extraordinary performance in classification tasks and generative solutions, neural networks have been applied to the spectral reconstruction problem. 
 
 This proceedings article will specifically discuss two recent proposals to improve on the extraction of transport coefficients from the Euclidean correlator via spectral functions. One is a particular neural network approach, mentioned above, and the other is a newly developed method that we have dubbed the multipoint method. This discussion serves as a starting point for the comparative study of systematic effects in several different spectral reconstruction techniques. The improved and newly proposed methods will firstly be tested on mock data to verify the validity of the predictions with spectral functions that reflect typical lattice QCD spectral functions. In the context of transport coefficients, the focus is on the extraction of $\rho (\omega) / \omega$ in the limit $\omega\to0$, which is typically directly related to the transport coefficients using Kubo formulas. For a number of other applications, at vanishing temperature in particular, it is either the full spectral function which is of relevance, or the spectral function at given frequencies. In this proceedings article, we focus on a particular Breit-Wigner peak for brevity and consider mostly the performance of the reconstruction techniques at vanishing and small frequencies. We leave a broader and more detailed comparison of the individual methods for a future publication. Finally, we give a brief outlook on the computation of the electric conductivity in quenched lattice QCD using different methods, including the two newly proposed ones.
\section{Spectral reconstruction}
Consider a correlator $G (\tau)$ measured on the lattice. Then this correlator is related to the spectral function $\rho (\omega)$ via the integral equation
\begin{equation}
    G(\tau) = \int_{0}^{\infty} \text{d} \omega \,  K(\tau, \omega) \rho (\omega) \, .
\label{eq:correlator integral equation}
\end{equation}
For the extraction of the spectral function, this integral equation needs to be inverted, which renders it an ill-posed problem due to the limited number of data points for the correlator $G (\tau)$ with finite accuracy.
The additional term in the integrand, $K(\tau,\omega)$, is the so-called kernel. For zero temperature, the kernel takes the form $K(\tau,\omega) = e^{-\omega \tau}$. Usually, finite $N_\tau$ corrections, present in realistic lattice simulations, are introduced through the symmetrized kernel 
\begin{equation}
    K(\tau, \omega) = e^{-\omega \tau} + e^{-\omega(1/T - \tau)}\,,
\label{eq:zero t kernel symm}
\end{equation}
where $1/T=a N_\tau$ signifies the time extent of the lattice on which the correlator $G (\tau)$ is measured. One may notice that \eqref{eq:zero t kernel symm} indeed becomes the non-symmetric zero temperature kernel in the limit $N_\tau \rightarrow \infty$.
The finite temperature kernel is defined as
\begin{equation}
    K(\tau, \omega) = \frac{\cosh [\omega (\tau - 1/2T)]}{\sinh (\omega/2T)}\,.
\label{eq:finite t kernel}
\end{equation}
Here, one may note that this kernel is indeed periodic, which reflects the reason why the symmetrized zero-temperature kernel is chosen.
\subsection{Unsupervised machine learning}
\label{sec:un-ml}
Machine-learning-based algorithms have recently gained attention in addressing the task of spectral function reconstruction \cite{Wang:2021jou, Buzzicotti:2023qdv, Kades:2019wtd}.  In this work, we adapt and use a previously developed "unsupervised" learning method~\cite{Wang:2021cqw,Wang:2021jou,Wang:2023fhc}. In this approach, the network learns a particular spectral density using a physics-informed loss function and the corresponding correlators. The authors originally proposed two architectures \cite{Wang:2021cqw} for this reconstruction, which they named NN and NN-P2P. The two architectures differ in their ability to learn the correlations between $\rho(\omega)$ at adjacent $\omega$-points.

\begin{figure}
    \centering
    \includegraphics[width=0.25\textwidth]{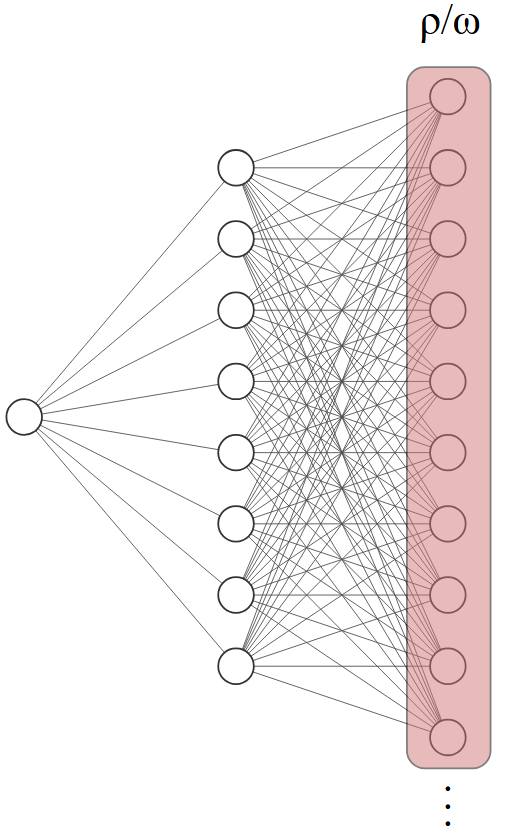} \hfill
    \includegraphics[width=0.6\textwidth]{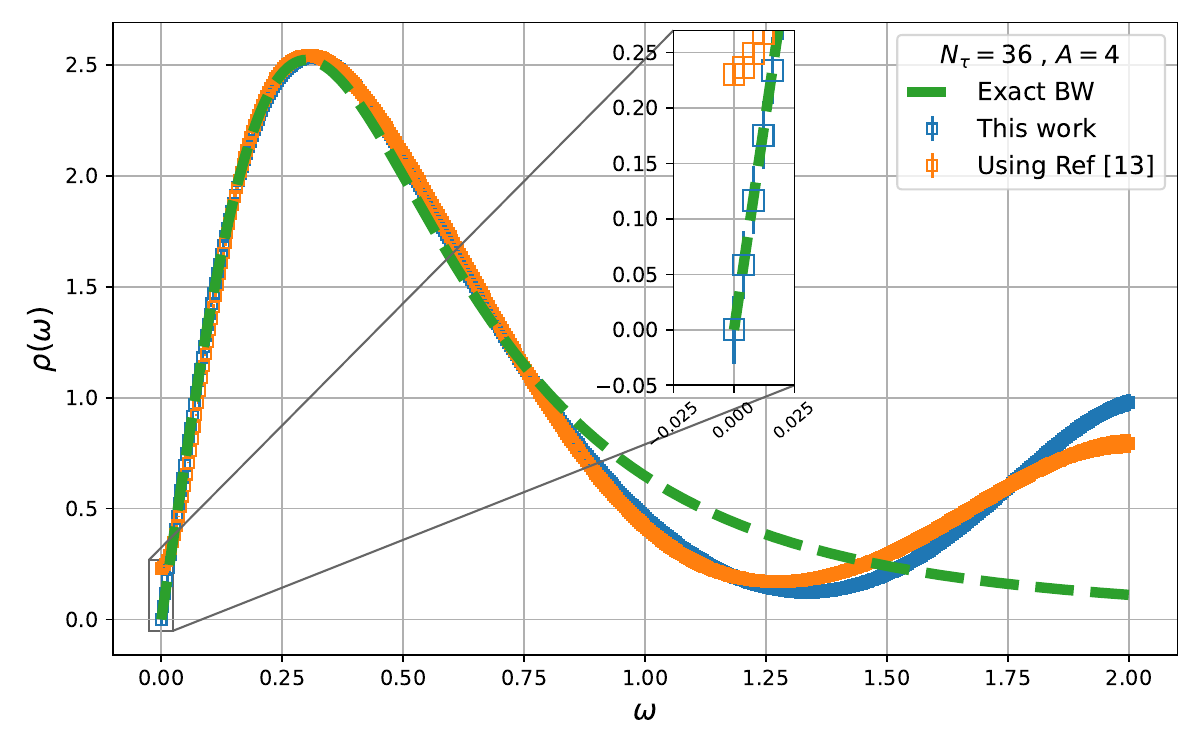}
    \caption{{\it Left panel:} Sketch of the unsupervised neural network structure. The shaded area signifies a softplus activation function. {\it Right panel:} Reconstruction of $\rho(\omega)$ as in~\cite{Wang:2021cqw} versus the direct extraction of $\rho(\omega)/\omega$ for a mock Breit-Wigner spectral function as described in Sec.~\ref{Mock data}.}
    \label{fig:comparison rho v rho/w training unsupervised}
\end{figure}

In this work, we use the NN version as shown in the left panel of Fig.~\ref{fig:comparison rho v rho/w training unsupervised}. In this case, the neural network is constructed with a constant input node $a^0 = C$ , $L$ hidden layers with $N_L$ nodes, and a final output layer with $N_\omega$ output nodes, representing $\rho(\omega)$ at $N_\omega$ discrete $\omega$ points used for the spectral function. We follow the original implementation by using the ELU activation function on the hidden layers and the Softplus activation function on the output layer, to ensure positivity of the spectral function. The main modification we made compared to the originally proposed algorithm was to train for $\rho(\omega)/\omega$, instead of $\rho(\omega)$. The immediate advantage of this is to recover $\rho(\omega)|_{\omega=0}=0$, and an improved sensitivity to the intercept $\rho(\omega)/\omega|_{\omega=0}$, which is required for the electric conductivity. A direct comparison of training for $\rho(\omega)$ versus training for $\rho(\omega)/\omega$ is shown for mock data in Fig:~\ref{fig:comparison rho v rho/w training unsupervised}.

For any machine learning task it is necessary to implement a loss function which parametrizes the accuracy of the prediction made by the network. For our task, the main contribution to the loss is the  correlator loss $L_{\rm corr}$ for the data $G_d$ and the error $\sigma$ on each correlator point
\begin{equation}
    L_{\rm corr} = \sum_{i=0}^{N_\tau} \frac{1}{\sigma_i^2} (G(\tau_i) - G_d(\tau_i))^2    \, .
    \label{eq:correlator loss}
\end{equation}
We implicitly assume here that the covariance matrix between the two correlators is given by $C (\tau_i, \tau_j) = \delta_{ij} 1/\sigma_i^2$. For this article, we will stick with this assumption since it is also beneficial for the accuracy of the training. As well as in~\cite{Wang:2023fhc}, two more terms have been added to the loss function to accommodate more controlled training, better convergence, and higher accuracy of the trained network, as well as a physical spectral function. Firstly, a $L2$ loss function $L_{L2}$ on the weights of the network, which serves as a regularizing tool against high fluctuations in the predictions of the network.
And secondly, a smoothness loss $L_{\rm smooth}$ on the predicted spectral function $\rho_{\rm pred}$ is added to the full loss function.
Thus, the full loss function $L$ for the neural network equates to
\begin{equation}
    L = L_{\rm corr} + \alpha L_{L2} + \beta L_{\rm smooth} \, .
    \label{eq:full loss}
\end{equation}
The scale setting parameters $\alpha$ and $\beta$ have been included to modify the weight of the contribution of the additional loss functions. These parameters can be tuned with libraries like ray-tune~\cite{liaw2018tuneresearchplatformdistributed} (which we have also tested previously) or can be set from experience, as done in this study.
\subsection{Multipoint method}
In this section, we discuss the novel method that we propose in these proceedings: the multipoint method. The aim is for the direct study of the slope of the spectral function in the limit $\omega\rightarrow 0$, which is related to transport coefficients via Kubo formulas for a multitude of spectral functions (see e.g~ Ref.~\cite{Meyer_2011} for a review). A particularly simple method to study the slope is to use the middle point of the Euclidean correlator~\cite{Aarts_2005}, which we will refer to as the midpoint method. In the following, we briefly discuss the midpoint method to prepare the ground for our extension to the multipoint method.

To understand why the midpoint provides a useful proxy for the slope of the spectral function at $\omega=0$, let us consider the spectral representation~\eqref{eq:correlator integral equation} with a finite temperature kernel. Assuming that $\rho(\omega)$ is an analytic function around $\omega=0$, we can expand it around that point, and the remaining integrals can be computed analytically. The result is given by
\begin{equation}
G(\tau = \alpha/T) = \sum_{n}\frac{c_n}{n!}\int_0^{\infty}\dd\omega\frac{\cosh[\omega(\tau-1/2T)]}{\sinh(\omega/2T)}\omega^n = \sum_{n}c_nT^{n+1}[\zeta(n+1,1-\alpha) + \zeta(n+1,\alpha)]
\label{eq:taylor_expansion_specf}
\end{equation}
where we introduced $\alpha \equiv \tau T$, with $0<\alpha<1/N_{\tau}$\footnote{\label{note:tau=0_comment} Notice that we excluded the $\alpha=0$ point since the Taylor series does not converge there.}, and $\zeta$ is the Hurwitz zeta function. Evaluating this expression at the midpoint, i.e.\ $\alpha=1/2$, we get
\begin{equation}
G(\tau=1/2T) = 6T^2\zeta_R(2)c_1 + 14 T^3\zeta_R(3) c_2 + 30T^4\zeta_R(4) c_3  + ...
\end{equation}
where we expressed the Hurwitz zeta functions $\zeta$ in terms of the Riemann zeta functions $\zeta_R$. We note that the variable change $\alpha = \tau T$ allows us to relate the Taylor expansion in $\omega$ to a similar expansion in $T$. Therefore, in the limit of $T\to0$, the following approximation holds
\begin{equation}
c_1 = \frac{1}{6T^2}G(\tau=1/2T) + \mathcal{O}(T)
\label{eq:midpoint_equation}
\end{equation}
In the $T\to0$ limit, the $\mathcal{O}(T)$ terms on the right-hand side vanish and Eq.~\eqref{eq:midpoint_equation} becomes exact. Hence, when $T$ is small, the midpoint offers an estimate of the slope $c_1 = [\rho(\omega)/\omega]_{\omega=0}$ of the spectral function at low frequency, and therefore of the value of the transport coefficient. An intuitive interpretation is offered when noticing that the combination $\omega K(\tau=1/2T,\omega)$ behaves like a Dirac delta for $T\to0$, thus giving out $\rho(\omega)/\omega$ at $\omega=0$. This method has already been applied to the electric conductivity~\cite{Buividovich:2020dks} and the out-of-equilibrium Chiral Magnetic Effect~\cite{Buividovich:2024bmu, Brandt:2025now}.

This relation, and with it the midpoint method, however, becomes inaccurate at higher temperatures -- the usual realm of interest for transport coefficients -- due to the higher-order terms in Eq.~\eqref{eq:midpoint_equation}. To systematically improve on the result of the midpoint method, one can use the remaining points of the correlator, since they too are related to the slope (c.f.\ Eq.~\eqref{eq:taylor_expansion_specf}). For a correlator with $N_{\tau}$ temporal points, one can use the following set of equations
\begin{align}
G(N_{\tau}/2) &= c_1A_{11} + c_2A_{12} + ... + c_{n_{\rm max}}A_{1 n_{\rm max}}  \nonumber\\
G(N_{\tau}/2-1) &= c_1A_{21} + c_2A_{22} + ... + c_{n_{\rm max}}A_{2n_{\rm max}} \nonumber\\
&... \nonumber\\
G(N_{\tau}/2-j) &= c_1A_{j+1,1} + c_2A_{j+1,2} + ... + c_{n_{\rm max}}A_{j+1, n_{\rm max}} \nonumber\\
&... \nonumber\\
G(N_{\tau}/2-(n_{\rm max}-1)) &= c_1A_{n_{\rm max}1} + c_2A_{n_{\rm max} 2} + ... + c_{n_{\rm max}}A_{n_{\rm max} n_{\rm max}}
\label{eq:multiple_correlator_eqs}
\end{align}
where $n_{\rm max} < N_{\tau}/2$. At $T \neq 0$, the coefficients $A_{kn}$ are given by 
\begin{equation}
A_{kn} = \frac{1}{N_{\tau}^{n+1}}\qty[\zeta(n+1,1-\alpha_k) + \zeta(n+1,\alpha_k)]\,,\hspace{0.5cm}\alpha_k = \frac{1}{2}-\frac{k-1}{N_{\tau}}\,.
\end{equation}
Thus the slope $c_1$ can be obtained more accurately by solving the $n_{\rm max} \times n_{\rm max}$ linear system of equations which can also be expressed as
\begin{equation}
\mqty(A_{11} & A_{12} &  ... & A_{1 n_{\rm max}} \\ A_{21} & A_{22} &  ... & A_{2 n_{\rm max}} \\ \vdots & \ddots & \ddots  & \vdots \\
A_{n_{\rm max}1} & A_{n_{\rm max}2} &  ... & A_{n_{\rm max} n_{\rm max}})\mqty(c_1 \\ c_2 \\ \vdots \\ c_{n_{\rm max}}) = \mqty(G(N_{\tau}/2) \\ G(N_{\tau}/2-1) \\ \vdots \\ G(N_{\tau}/2-(n_{\rm max}-1))) \, .
\label{eq:multipoint_linear_system}
\end{equation}
This is the multipoint method. The central point of this novel approach is to extend the midpoint method by using all the information available from the remaining points of the correlator, excluding $\tau=0$ (c.f.\ footnote~\ref{note:tau=0_comment}). The set of equations~\eqref{eq:multiple_correlator_eqs} allows one to determine $c_1$ by canceling higher-order terms in the Taylor expansion, thus systematically improving the approximation in the finite-$T$ case. Note that, for this article, only bosonic correlators will be of interest. This constrains the linear system of equations to odd powers in $\omega$ in the Taylor series (see e.g.\ Ref.~\cite{Meyer_2011}). For this special case, a similar slightly modified linear system can be constructed.

The multipoint method is comparable to the well-established Backus-Gilbert (BG) method since  \begin{wrapfigure}{r}{0.5\textwidth}
\centering
\includegraphics[width=.9\linewidth]{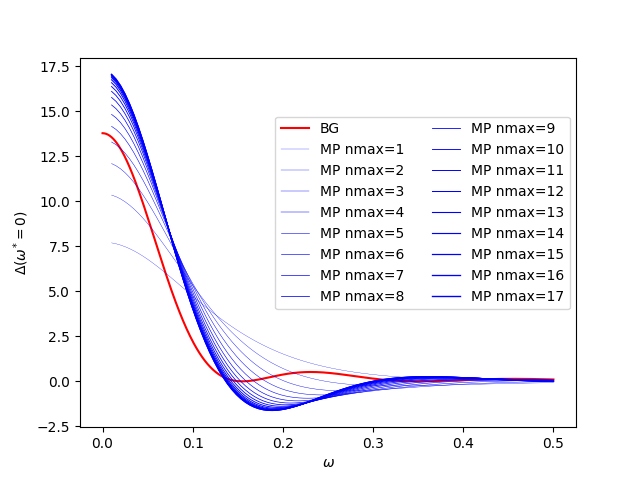}
\caption{Comparison of the smearing kernels with $N_t=36$ obtained from the unregularized BG method and the multipoint (MP) with different numbers of used correlator points.}
\label{fig:multi-kernel}
\end{wrapfigure}
both correspond to a linear combination of the correlator points. In Fig.~\ref{fig:multi-kernel} we compare the smearing kernel of both procedures at $\omega=0$, which serves as a proxy for how reliably these methods can extract $\rho(\omega)/\omega$ when $\omega\rightarrow 0$. We see that both kernels are similarly shaped for a large number of points included for the multipoint method, maybe with a slight advantage for the BG method. Note that both methods need an inversion of a badly conditioned matrix. This can be done similarly in both cases, and the results, in particular, smearing kernels, should be compared after regularization. We will perform this comparison and a more detailed presentation of the differences between the two methods in an upcoming publication. 
\section{Mock data}
\label{Mock data}
Systematically analysing the performance of different spectral reconstruction methods needs an extensive testing procedure where knowledge of the spectral function is available. Thus, we turn to mock data before applying the methods to lattice data. In this proceedings article, we will focus on a mock data test using a simple Breit-Wigner peak as an example, which has connections to what one would expect for the spectral function associated with the electric conductivity. The spectral function that is being reconstructed in this chapter can be parameterized as 
\begin{equation}
    \rho(\omega) = \frac{4\alpha g\omega}{4g^2\omega^2 + [m^2 + g^2 - \omega^2]^2}\,,\hspace{0.5cm}\text{where}\hspace{0.5cm}\begin{array}{cc}
    \tilde{\alpha} = a^2\alpha &= 0.5 \\
    \tilde{m} = a m &= 0.1 \\
    \tilde{g} = a g &= 0.5
    \end{array}
    \label{eq:breit wigner spectral function}
\end{equation}
with the parameters being fixed in lattice units. The choice of parameters is, in principle, arbitrary, but for simplicity, we have chosen them such that $\rho$ is dimensionless. 

In this proceedings article, we will focus on the simplest case of uncorrelated noise for the mock correlator. Correlated and more realistic mock data will be discussed in an upcoming publication. For the noise generation of the correlation function, a similar strategy to~\cite{Asakawa_2001} has been applied. For a specific choice of the Kernel $K(\omega, \tau)$ the exact correlator for \eqref{eq:breit wigner spectral function} can be calculated by
\begin{equation}
    G_{\rm exact} (\tau) = \int_0^\infty \dd\omega K(\omega, \tau) \rho (\omega) \, .
\end{equation}
By sampling from a Gaussian distribution $\mathbf{N}(0,\sigma (\tau, A))$ for different noise levels $A \in \{2,3,4\}$ with 
\begin{equation}
    \sigma (\tau, A) = 10^{-A} G_{\rm exact} (\tau) \frac{(\tau +\Delta\tau)}{\Delta\tau}\,,
\end{equation}
where $\Delta \tau$ is the lattice spacing, the noise for the exact correlator is obtained. Defining the mock correlator for a specific noise level as
\begin{equation}
    G^A_{\rm mock}(\tau) = G_{\rm exact} (\tau) + \eta (\tau, A) \, ,
\end{equation}
where $\eta (\tau, A)$ is the Gaussian random number at $\tau$ and $A$.
\section{Results}
In this section, we will present our results obtained using several different spectral reconstruction techniques. First, we apply the techniques to the mock data. Secondly, we aim at extracting the slope of the spectral function of the current-current correlators, and with it the electric conductivity, at non-zero external magnetic fields in longitudinal and transversal directions. The associated correlation function has been evaluated in the quenched approximation at a temperature of 1.45$T_c$.
\subsection{Mock data}
\begin{figure}
    \centering
    \includegraphics[width=\textwidth]{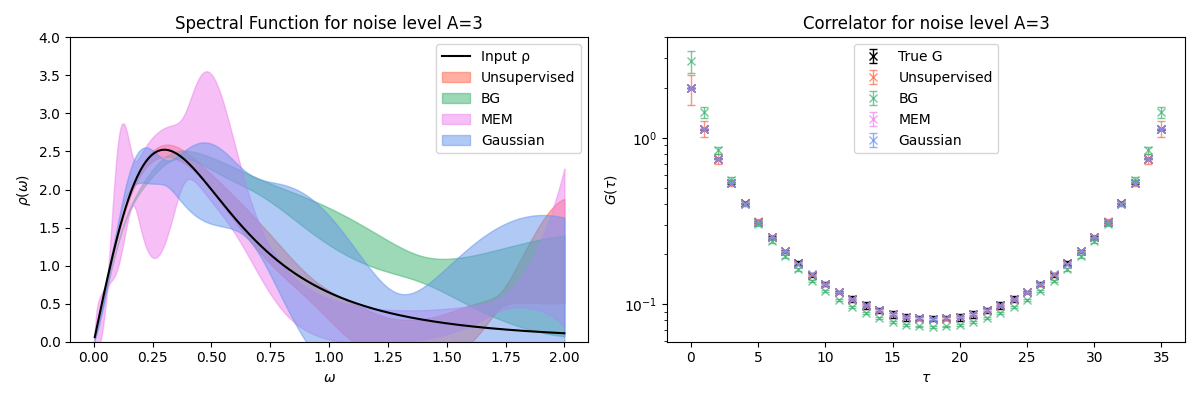} \\[-1mm]
    \includegraphics[width=\textwidth]{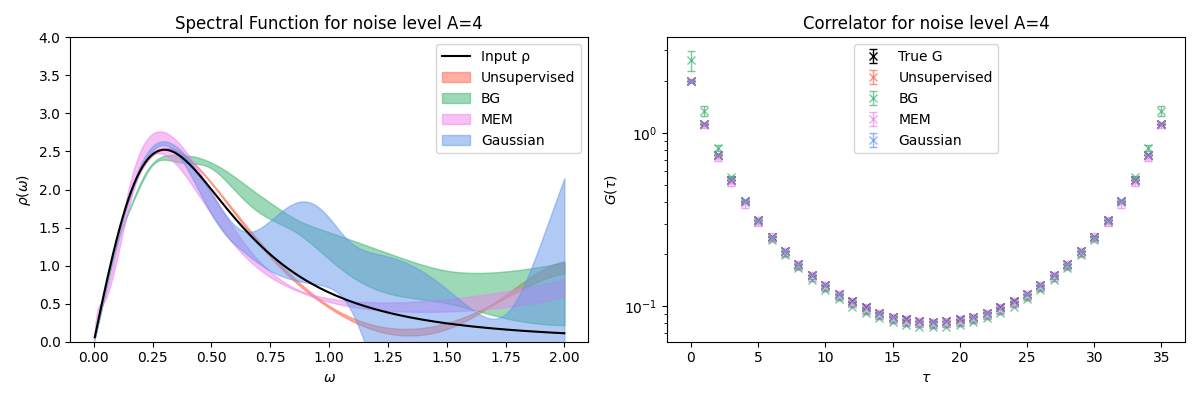}
    \caption{Comparison of the performance of the machine learning method (``unsupervised''), the Gaussian method, BG, and MEM on the mock data described in Sec.~\ref{Mock data} with $N_\tau = 36$ and a noise level of $A=3$ (top) and $A=4$ (bottom).}
    \label{fig:mock_36_noise3}
\end{figure}

For the previously described mock data, the Gaussian method, the BG method, MEM, as well as the machine learning framework from Sec.~\ref{sec:un-ml} (unsupervised) have been used to reconstruct the spectral function $\rho (\omega)$ from \eqref{eq:breit wigner spectral function}. Figure~\ref{fig:mock_36_noise3} shows the reconstructions in terms of $\rho(\omega)$ with decreasing noise level in the input correlator, along with systematic errors arising from the methods themselves. As a comparison, the true spectral function is also shown, and the right panels also show the reconstructed correlators with statistical errors. All of the tested methods reproduce the slope of $\rho(\omega)$ at $\omega \rightarrow 0$ as well as the position of the peak with reasonable accuracy. At higher $\omega$, the predictions worsen significantly, which we will investigate and discuss further in future studies. The shape of the correlator is matched by almost all of the methods. The exception is the BG, which, however, is known to produce a strongly smeared spectral function that does not follow in all details the true spectral function.

\subsection{Lattice calculation}
To test the methods on actual lattice data, we have chosen to consider the electric conductivity coefficient, which is an important quantity for the modeling of peripheral heavy-ion collisions where magnetic fields appear.
For the lattice setup, we consider quenched QCD at non-zero temperature with O(a)-improved Wilson valence fermions with a varying number of B-fields to study the effect of different magnitudes of the electric conductivity coefficient on the spectral reconstruction. The configurations for the measurement of this operator are generated with the SU(3) plaquette action and the simulation parameters correspond to the $N_s=48, N_t=16$ lattice from Ref.~\cite{Ding_2011}.

\begin{figure}
    \centering
    \includegraphics[width=0.35\textwidth]{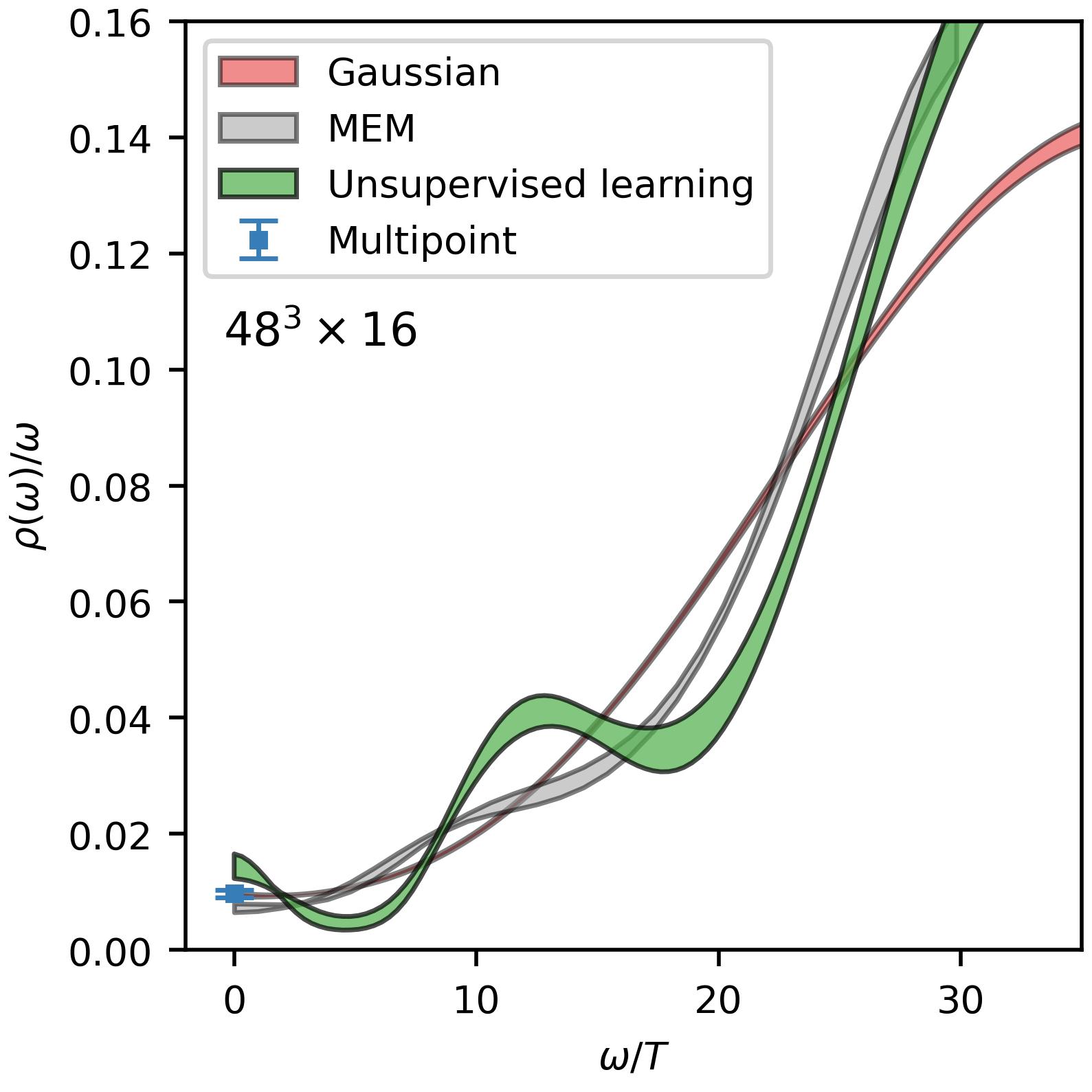}
    \includegraphics[width=0.35\textwidth]{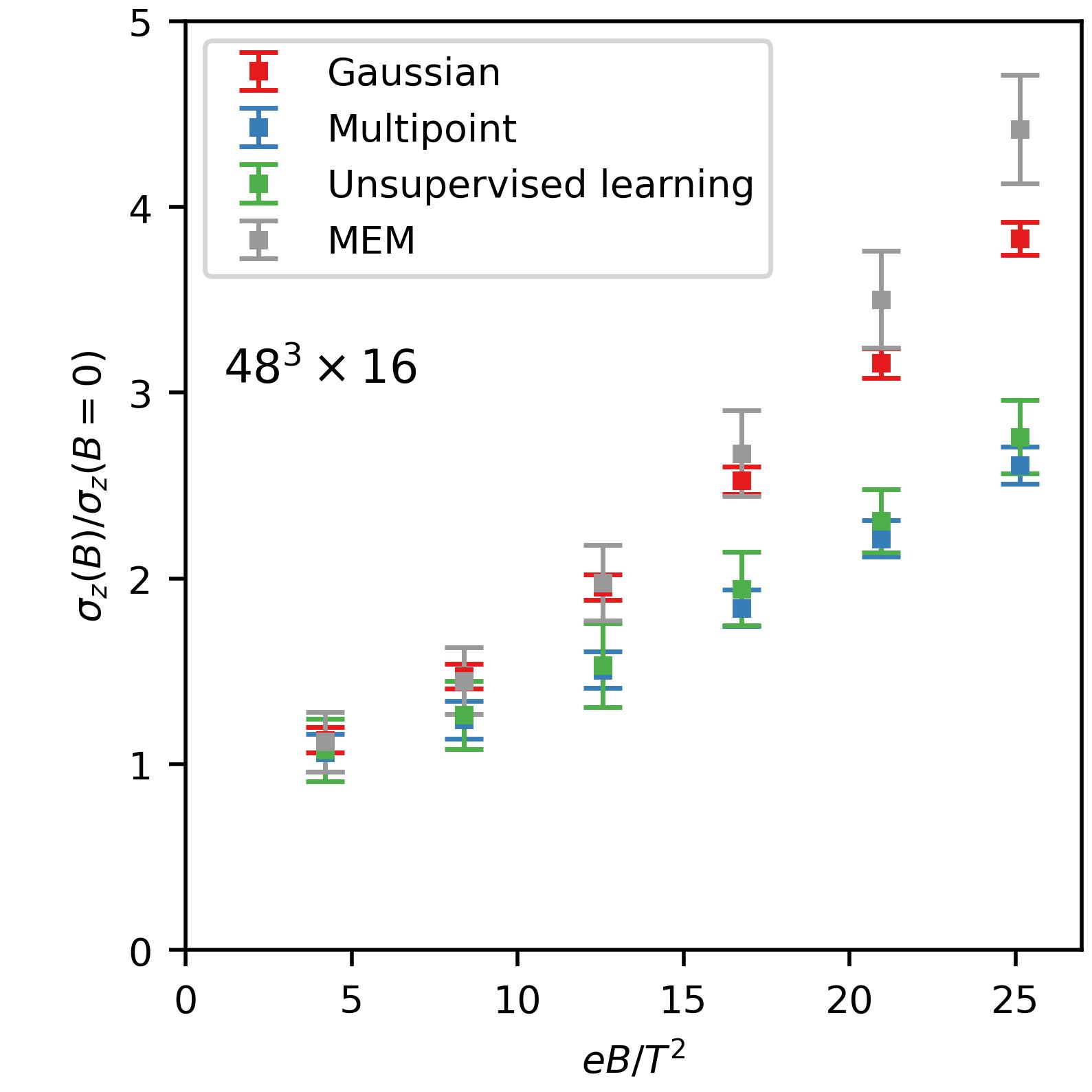}
    \caption{Left: $\rho_{33}(\omega)/\omega$, at $N_b=6$, obtained using various spectral reconstruction methods, including the Gaussian method via the fredipy package \cite{fredipy,Horak:2021syv}. Right: Longitudinal electric conductivity as a function of the magnetic field strength. Here, the error bars only include statistical errors. We observe qualitative agreement with results from full QCD staggered simulations~\cite{Astrakhantsev_2020}.}
    \label{fig:electric conds parallel}
\end{figure}

We consider the connected part of the local current-current correlator in the vector channel, and we concentrate  on the longitudinal direction, including the spectral function $\rho_{33}$, with respect to the magnetic field~ Ref.~\cite{Astrakhantsev_2020}.
From the measured lattice correlator, we then obtain the spectral function via spectral reconstruction. The results are shown in the left panel of Figure~\ref{fig:electric conds parallel}. Calculating the conductivity coefficient for this observable follows from the Kubo formula given in~\cite{Ding_2011} by
\begin{equation}
    \frac{1}{C_{em}} \frac{\sigma_z}{T} = \lim_{\omega \rightarrow 0}\frac{\rho_{33} (\omega)}{6 \omega} = \mathcal{C} \, .
\label{eq:Kubo formula electric cond coeff}
\end{equation}
The results are shown on the right-hand side of Figure~\ref{fig:electric conds parallel} as a function of the magnetic field strength.

\hspace*{-5mm}
\section{Conclusions}
In this work, we studied several different spectral reconstruction techniques using mock data. In particular, this includes an adaptation of the unsupervised machine learning approach previously suggested in~\cite{Wang:2023fhc} as well as a newly developed multipoint method for the determination of the conductivity coefficient, building on the basis of the proposed midpoint method in~\cite{Aarts_2005}. For this simple input spectral function, it has been found that the employed reconstruction techniques accurately reconstruct the mock spectral function, but develop large individual systematic uncertainties. Notably, all methods agree on the shape and the slope at $\omega \rightarrow 0$, relevant for the extraction of transport coefficients. The method that deviates most prominently from the input spectral function is the BG method, due to the smearing of the spectral function returned by the method.

Using lattice simulations, the longitudinal electric conductivity coefficient $\sigma_z$ has been studied. We have observed an increase of $\sigma_z$ with the field strength of the magnetic field, which is consistent between different reconstruction techniques, despite showing quantitative differences that deserve further study. Furthermore, the result is in qualitative agreement with earlier full QCD staggered results~\cite{Astrakhantsev_2020}.
Further studies will be discussed in a future publication.
\hspace*{-5mm}
\acknowledgments
The authors thank David Clarke for support with the MEM implementation. This work was supported by the Deutsche Forschungsgemeinschaft (DFG, German Research Foundation) – project number 315477589 – TRR 211 as well as
the Helmholtz Graduate School for Hadron and Ion Research (HGS-HIRe for FAIR). SS gratefully acknowledges access to the Marvin cluster of the University of Bonn and is supported by the Deutsche Forschungsgemeinschaft (DFG, German Research Foundation) as part of the CRC 1639 NuMeriQS -- project no.\ 511713970. It also received support from MICIU/AEI/10.13039/501100011033 and FEDER (EU) under Grant PID2022-140440NB-C21 and by Consejeria de Universidad, Investigación y Innovación and Gobierno de España and EU -- NextGenerationEU, under Grant AST22 8.4. EGV acknowledges funding from the Hungarian National Research, Development and Innovation Office (Research Grant Hungary 150241) and the European Research Council (Consolidator Grant 101125637 CoStaMM).
\hspace*{-3mm}

\begin{thebibliography}{99}
\bibitem{Bryan_MEM}
R.~K.~Bryan, 
Eur. Biophys. J \textbf{18} (1990), 165–174
%
\bibitem{Asakawa_2001}
M.~Asakawa, T.~Hatsuda and Y.~Nakahara,
Prog. Part. Nucl. Phys. \textbf{46} (2001), 459-508
[arXiv:hep-lat/0011040 [hep-lat]].
%
\bibitem{Ding_2011}
H.~T.~Ding, A.~Francis, O.~Kaczmarek, F.~Karsch, E.~Laermann and W.~Soeldner,
Phys. Rev. D \textbf{83} (2011), 034504
[arXiv:1012.4963 [hep-lat]].
%
\bibitem{Meyer_2011}
H.~B.~Meyer,
Eur. Phys. J. A \textbf{47} (2011), 86
[arXiv:1104.3708 [hep-lat]].
%
\bibitem{Ding_2019}
H.~T.~Ding, O.~Kaczmarek, A.~L.~Kruse, R.~Larsen, L.~Mazur, S.~Mukherjee, H.~Ohno, H.~Sandmeyer and H.~T.~Shu,
Nucl. Phys. A \textbf{982} (2019), 715-718
[arXiv:1807.06315 [hep-lat]].
%
\bibitem{Horak_2022}
J.~Horak, J.~M.~Pawlowski, J.~Rodr{\'\i}guez-Quintero, J.~Turnwald, J.~M.~Urban, N.~Wink and S.~Zafeiropoulos,
Phys. Rev. D \textbf{105} (2022) no.3, 036014
[arXiv:2107.13464 [hep-ph]].
%
\bibitem{Hansen:2019idp}
M.~Hansen, A.~Lupo and N.~Tantalo,
Phys. Rev. D \textbf{99} (2019) no.9, 094508
[arXiv:1903.06476 [hep-lat]].
%
\bibitem{BG1968}
G.~Backus and F.~Gilbert,
Geophys. J. Int. \textbf{16} (1968) no.2, 169-205
%
\bibitem{Wang:2023fhc}
L.~Wang, S.~Shi and K.~Zhou,
J. Phys. Conf. Ser. \textbf{2586} (2023) no.1, 012158
%
\bibitem{Kades:2019wtd}
L.~Kades, J.~M.~Pawlowski, A.~Rothkopf, M.~Scherzer, J.~M.~Urban, S.~J.~Wetzel, N.~Wink and F.~P.~G.~Ziegler,
Phys. Rev. D \textbf{102} (2020) no.9, 096001
[arXiv:1905.04305 [physics.comp-ph]].
%
\bibitem{Wang:2021jou}
L.~Wang, S.~Shi and K.~Zhou,
Phys. Rev. D \textbf{106} (2022) no.5, L051502
[arXiv:2111.14760 [hep-ph]].
%
\bibitem{Buzzicotti:2023qdv}
M.~Buzzicotti, A.~De Santis and N.~Tantalo,
Eur. Phys. J. C \textbf{84} (2024) no.1, 32
[arXiv:2307.00808 [hep-lat]].
%
\bibitem{Wang:2021cqw}
L.~Wang, S.~Shi and K.~Zhou,
[arXiv:2112.06206 [physics.comp-ph]].
%
\bibitem{liaw2018tuneresearchplatformdistributed}
R.~Liaw, E.~Liang, R.~Nishihara, P.~Moritz, J.~E.~Gonzalez and I.~Stoica,
[arXiv:1807.05118 [cs.LG]].
%
\bibitem{Aarts_2005}
G.~Aarts and J.~M.~Martinez Resco,
Nucl. Phys. B \textbf{726} (2005), 93-108
[arXiv:hep-lat/0507004 [hep-lat]].
%
\bibitem{Buividovich:2020dks}
P.~V.~Buividovich, D.~Smith and L.~von Smekal,
Phys. Rev. D \textbf{102} (2020) no.9, 094510
[arXiv:2007.05639 [hep-lat]].
%
\bibitem{Buividovich:2024bmu}
P.~V.~Buividovich,
Phys. Rev. D \textbf{110} (2024) no.9, 094508
[arXiv:2404.14263 [hep-lat]].
%
\bibitem{Brandt:2025now}
B.~B.~Brandt, G.~Endrodi, E.~Garnacho Velasco, G.~Marko and A.~D.~M.~Valois,
PoS \textbf{LATTICE2024} (2025), 196
[arXiv:2502.01155 [hep-lat]].
%
\bibitem{fredipy}
J.~Turnwald., J.~M.~Urban,
[github:JonasTurnwald/fredipy].
%
\bibitem{Horak:2021syv}
J.~Horak, J.~M.~Pawlowski, J.~Rodr{\'\i}guez-Quintero, J.~Turnwald, J.~M.~Urban, N.~Wink and S.~Zafeiropoulos,
Phys. Rev. D \textbf{105} (2022) no.3, 036014
[arXiv:2107.13464 [hep-ph]].
%
\bibitem{Astrakhantsev_2020}
N.~Astrakhantsev, V.~V.~Braguta, M.~D'Elia, A.~Y.~Kotov, A.~A.~Nikolaev and F.~Sanfilippo,
Phys. Rev. D \textbf{102} (2020) no.5, 054516
[arXiv:1910.08516 [hep-lat]].
%
\end{thebibliography}
{}
\end{document}